# Spin/momentum properties of the paraxial optical beams

Peng Shi, Heng Li, Luping Du*, Xiaocong Yuan*

Institute of Micro/Nano Optoelectronics, Shenzhen University, 518060, China

*Authors to whom correspondence should be addressed: *lpdu@szu.edu.cn*, and *xcyuan@szu.edu.cn*

**Abstract:** Spin angular momentum, an elementary dynamical property of classical electromagnetic fields, plays an important role in spin–orbit and light–matter interactions, especially in near-field optics. The research on optical spins has led to the discovery of phenomena such as optical spin–momentum locking and photonic topological quasiparticles, as well as applications in high-precision detection and nanometrology. Here, we investigate spin–momentum relations in paraxial optical systems and show that the optical spin angular momentum contains transverse and longitudinal spin components simultaneously. The transverse spin originates from inhomogeneities of field and governed by the vorticity of the kinetic momentum density, whereas the longitudinal spin parallel to the local canonical momentum is proportional to the polarization ellipticity of light. Moreover, the skyrmion-like spin textures arise from the optical transverse spin can be observed in paraxial beams, and their topologies are maintained free from the influence of the Gouy phase during propagation. Interestingly, the optical singularities, including both phase and polarization singularities, can also affect the spin–momentum properties significantly. Our findings describe the intrinsic spin–momentum properties in paraxial optical systems and apply in the analysis of the properties of spin–momentum in optical focusing, imaging, and scattering systems.

## I. Introduction

Spin angular momentum (SAM) is a fundamental dynamical property of elementary particles and classical wave fields[1–14] and plays a critical role in understanding spin–orbit[15–29] and wave–matter[1,2,30–32] interactions and predicting the behavior of interacting systems. For a classical electromagnetic (EM) field, the SAM associated with circular or elliptical polarization can be orientated in an arbitrary direction[15,16,33,34]. From this general perspective, for plane-wave solutions of Maxwell's equations, the SAM component oriented along the propagating direction (defined by the wavevector or canonical momentum) is considered as the longitudinal spin[33] (L-spin), whereas the SAM component oriented perpendicular to the wavevector is the transverse spin[34] (T-spin). To date, T-spins have been intensively investigated in various EM systems, including confined optical systems such as evanescent waves[35–40], guided modes[41–45], and free-space optical systems such as the Gaussian focused fields[46–50], interference fields[51], nondiffracting fields[52], and unpolarized fields[53–56]. Remarkably, in confined optical systems, the T-spins of evanescent and guided modes feature a well-known property referred to as spin–momentum locking[36,41,42,57,58] and their spin–orbit couplings raise a large class of remarkable phenomena, such as unidirectional guided wave[59–67] and photonic topological quasiparticles[68–78], and offer potential applications in angular-momentum-based optical manipulation[37,49,50,79,80], imaging[81–93], detection[94–96], metrology[97,98], and on-chip quantum technologies[99].

Previously, in free space, only the spin properties of special optical beams were investigated[46–52]. However, for an arbitrary structured light which carries the inhomogeneities of intensity, phase, polarization and singularities[100–105], a unified methodology is lacking in describing the dynamical evolving of momentum and angular momentum, especially regarding paraxial focusing, imaging, and scattering systems. Moreover, although optical beams in the free space can form skyrmionic beams[106–118], the Gouy phase governs the continuous evolution of polarization structures in the propagating beam, whereas SAM is a good candidate in describing the topological invariants of propagating optical beams. Therefore, understanding the spin–momentum properties of paraxial optical fields is meaningful in providing a guide for spin-state manipulation.

Here, we investigate the spin–momentum properties of paraxial optical systems and present a unified methodology to perform the decomposition of the optical spin into T-spin and L-spin for various types of paraxial optical beams. The results reveal that there are optical T-spins that govern optical spin–momentum locking and L-spins determined by the polarization ellipticities (helicities) that are parallel to the canonical momentum in paraxial optical beams. Remarkably, from the spin–momentum locking derived from optical T-spin, skyrmion-like spin textures can form in paraxial optical beams and their topologies are maintained during propagation, free from the influence of the Gouy phase. We furthermore investigate the influence of optical singularities on the spin–momentum properties in paraxial optical systems and discover an extraordinary SAM component. The direction of its vector, which is perpendicular to that of canonical momentum, is not determined by the vorticity of the kinetic momentum but by the polarization topological charge of the vector vortex beam. This SAM component is characterized by $\mathbb{Z}_4$ topological invariants and should be considered as a L-spin. Our findings are general for high-order structured light beams constructed through the linear superpositions of paraxial optical modes and can help in the understanding of the dynamical properties of light and broadens the study of topological quasiparticles in paraxial systems.

## II Theory
**General theoretical results**

For monochromatic, time-harmonic EM waves in the paraxial approximation with complex electric field strength $\mathbf{E}$ and magnetic field strength $\mathbf{H}$ having angular frequency dependence $\omega$, we demonstrate that the kinetic momentum $\mathbf{p} = \text{Re}\{\mathbf{E}^* \times \mathbf{H}\}/2c^2$, with $c$ the velocity of light in vacuo, and the SAM $\mathbf{S} = \text{Im}\{\varepsilon(\mathbf{E}^* \times \mathbf{E}) + \mu(\mathbf{H}^* \times \mathbf{H})\}/4\omega$, with $\varepsilon$ and $\mu$ the permittivity and permeability of the vacuum and superscript $*$ signifying complex conjugation[119–123], have an inherent relationship given by

$$\mathbf{S}_T = \frac{1}{2k^2}\nabla \times \mathbf{p} \quad \text{and} \quad \mathbf{S}_L = \mathbf{S} - \mathbf{S}_T. \tag{1}$$

Here, $k = \omega/c$ denotes the wave number, $\mathbf{S}_T$ and $\mathbf{S}_L$ denote the optical T-spin and L-spin, respectively. These equations show that, in the paraxial optical systems, the T-spin stems from the field inhomogeneities and given by the vorticity of the kinetic momentum density. The transversality of the optical T-spin is confirmed from the identity $\nabla \cdot (\nabla \times \mathbf{A}) = 0$. However, it does not mean that the curl of the kinetic momentum is always perpendicular to the kinetic momentum. By decomposing the structured light into a superposition of plane waves, the perpendicularity of the momentum vector and the optical T-spin vector is satisfied for each plane wave basis, as for example the evanescent plane wave[35,36]. Remarkably, the T-spin is capable of producing spin–momentum locking[36,41,42,57], i.e., if the kinetic momentum is reversed, the T-spin is inverted correspondingly. In addition, the T-spin is a classical physical quantity, and the kinetic momentum describing the group velocity of photons is considered as the current of light[119,121]. Thus, spin–momentum locking from the T-spin, which may also be considered as spin–current locking, originates from the intrinsic spin–orbit property inherent in Maxwell's equations and is different from the quantum spin Hall effect in condensed matter physics[124].

In contrast, the L-spin is extracted by subtracting the T-spin from the total SAM. In the following, we shall demonstrate that this difference, which is related to the three-dimensional (3D) polarization ellipticities along the local wavevector[15–17] and the Berry curvature of paraxial optical fields[15], is definitely the L-spin. Here, we primarily focus our attention on the spin–momentum properties of the paraxial modes, the solutions of the Helmholtz equation, in both Cartesian and cylindrical coordinates, and high-order structured light beams constructed from linear superpositions of these modes[125–130].

**Spin-momentum property of Hermite-Gaussian beams**

We first consider a Hermite-Gaussian (HG) beam, the solution to the Helmholtz equation in the paraxial approximation in Cartesian coordinates $(x, y, z)$ with unit vector $(\hat{x}, \hat{y}, \hat{z})$, propagating along the $z$ axis. The electric and magnetic field components can be expressed as

$$\mathbf{E}_{HG} = \left[ +\eta_x u_{HG}\hat{x}, +\eta_y u_{HG}\hat{y}, +\frac{1}{ik}\left(\eta_x \frac{\partial}{\partial x} + \eta_y \frac{\partial}{\partial y}\right) u_{HG}\hat{z} \right]^T e^{-ikz} \quad (2)$$

and

$$\mathbf{H}_{HG} = \frac{k}{\omega\mu}\left[ +\eta_y u_{HG}\hat{x}, -\eta_x u_{HG}\hat{y}, +\frac{1}{ik}\left(\eta_y \frac{\partial}{\partial x} - \eta_x \frac{\partial}{\partial y}\right) u_{HG}\hat{z} \right]^T e^{-ikz}, \quad (3)$$

where $\eta_x$ and $\eta_y$ are arbitrary complex constants describing the relative strength, $\text{Im}\{\eta_x^* \eta_y\}$ specifies the polarization ellipticity (helicity) of the paraxial HG beam, and the superscript T indicates the transpose of the matrix. The electric field $\mathbf{E}_{HG}$ and magnetic field $\mathbf{H}_{HG}$ also satisfy Gauss's law ($\nabla \cdot \mathbf{E}_{HG} = 0$ and $\nabla \cdot \mathbf{H}_{HG} = 0$) in the paraxial approximation ($\frac{\partial^2 u_{HG}}{\partial z^2} \ll k \frac{\partial u_{HG}}{\partial z} \ll k^2 u_{HG}$). The complex amplitude $u_{HG}$ is given by

$$u_{HG,mn} = \frac{w_0}{w(z)} H_m\left[\frac{\sqrt{2}x}{w(z)}\right] H_n\left[\frac{\sqrt{2}y}{w(z)}\right] \exp\left(-\frac{x^2+y^2}{w^2(z)} - i\frac{k(x^2+y^2)}{2R(z)}\right) \exp\left(-i(1+m+n)\tan^{-1}\left(\frac{z}{z_R}\right)\right). \quad (4)$$

Here, $H_m(x)$ is the Hermite polynomial with non-negative integer index $m$, $z_R = \pi w_0^2/\lambda$ the Rayleigh range, $w(z) = w_0\sqrt{1-z^2/z_R^2}$ the beam width of the propagating wave, $w_0$ the beam radius at the beam waist, $R(z)$ the radius of curvature of the wavefronts, $\lambda$ the wavelength, and the last factor $\exp(-i(1+m+n)\tan^{-1}(z/z_R))$ is the Gouy phase[107]. From Equation 2, 3, and 4, one finds that the paraxial HG beam displays both intensity and phase inhomogeneities, whereas the polarization is homogeneous in the transverse propagating plane ($xy$-plane).

Employing Equation 2 and 3, the kinetic momentum of paraxial HG beam is

$$\mathbf{p} = \frac{\varepsilon k}{2\omega} \text{Re} \begin{pmatrix} \frac{1}{ik}\left[+\left(\eta_x^*\eta_x + \eta_y^*\eta_y\right)u_{HG}^* \frac{\partial u_{HG}}{\partial x} + \eta_x^*\eta_y \frac{\partial u_{HG}^* u_{HG}}{\partial y}\right]\hat{x} \\ \frac{1}{ik}\left[+\left(\eta_x^*\eta_x + \eta_y^*\eta_y\right)u_{HG}^* \frac{\partial u_{HG}}{\partial y} - \eta_x^*\eta_y \frac{\partial u_{HG}^* u_{HG}}{\partial x}\right]\hat{y} \\ \frac{1}{ik}\left[-ik\left(\eta_x\eta_x^* + \eta_y\eta_y^*\right)u_{HG}^* u_{HG} - 0\right]\hat{z} \end{pmatrix}. \quad (5)$$

From the spin–orbit decomposition theory of the kinetic momentum for classical EM fields[41,42,119,121], the kinetic momentum of an optical field can be decomposed into canonical and spin momentum components ($\mathbf{p} = \mathbf{p}_o + \mathbf{p}_s$). One recognizes the first term in Equation 5 as the canonical momentum ($\mathbf{p}_o \propto \langle \psi | i \nabla | \psi \rangle$ with $\psi$ the 6-vector photonic wave function[131,132]), and the z-component of the canonical momentum as being proportional to the energy density ($(\eta_x^* \eta_x + \eta_y^* \eta_y) u_{HG}^* u_{HG}$). In contrast, the second term in Equation 5 is the Belinfante spin momentum $\mathbf{p}_s = \nabla \times \mathbf{S}/2$[119,121], in which the SAM is proportional to the polarization ellipticity $\text{Im}\{\eta_x^* \eta_y\} u_{HG}^* u_{HG}$[33]. Generally, for structured light beams, the Belinfante spin momentum is nonzero, and thus the kinetic momentum is not parallel to the canonical momentum, which determines the local wavevector $\mathbf{k}$ through relation $\mathbf{p}_o = \hbar \mathbf{k}$.

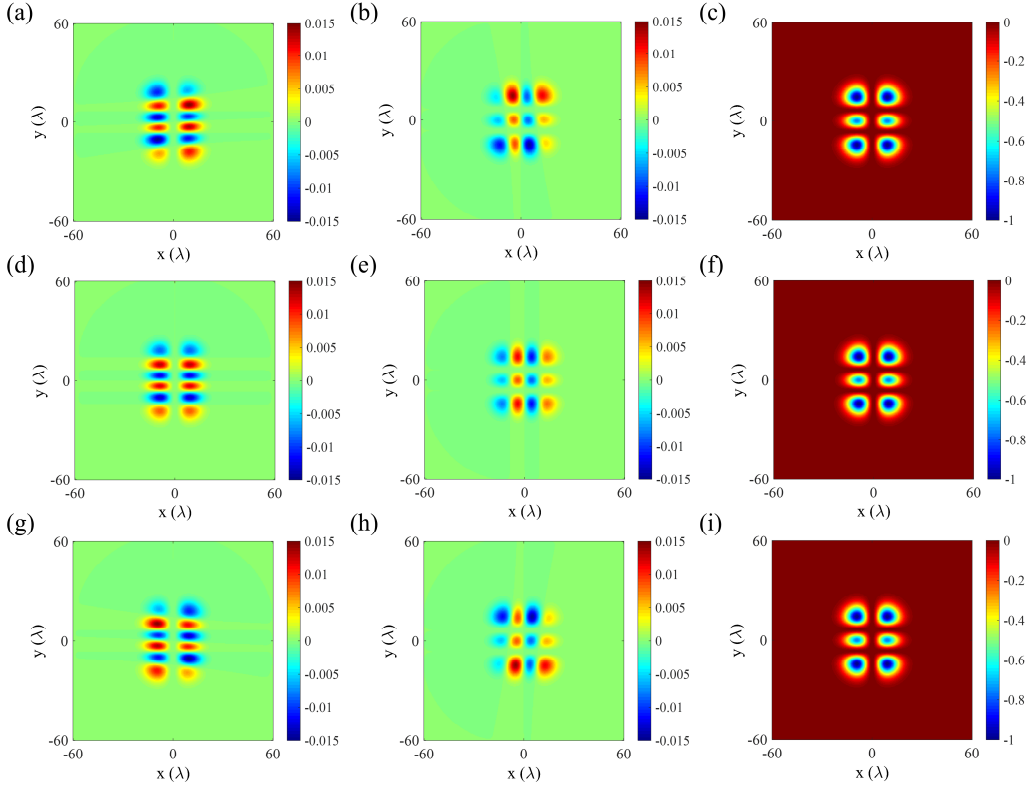

**Figure 1.** Momentum properties of paraxial $HG_{12}$ beams in the pre-focal, focal and post-focal planes: (a) x-component $p_x$, (b) y-component $p_y$, and (c) z-component $p_z$ of the kinetic momentum at plane $z = -100\lambda$. By analyzing their vector structure, the horizontal kinetic momentum components contain multiple vortex structures. The z-component kinetic momentum is always nonzero, hence the photons of this HG beam precess during propagation. (d–f) as for (a–c), but at plane $z = 0.001\lambda$; (g–i) same as (a–c), but at plane $z = +100\lambda$. Here, $\eta_x = 1$, $\eta_y = 2i$, $w_0 = 8$ μm, $\lambda = 0.6328$ μm.

To understand the momentum properties of paraxial HG beams, we plot the three components of the kinetic momentum densities for the $HG_{12}$ mode in the pre-focal plane, focal plane, and post-focal plane in Figure 1. From Figure 1a-b, d-e, and g-h, the horizontal kinetic momenta differ slightly through convergence in the pre-focal plane and divergence in the post-focal plane compared with those in the focal plane. By analyzing the vector properties of these horizontal kinetic momentum densities, one can find that they contain multiple vortex structures (Figure 3a). Moreover, from Figure 1c, f and i, one finds that the z-component kinetic momenta in these planes are parallel. Thus, the photons of this paraxial HG mode precess around the z-axis and these spiral trajectories of the photons give rise to the geometric phase[133-135]. Note that the Gouy phase in Equation 4 does not affect the properties of the kinetic momenta when the beam passes through the focal plane, and thus the vortex structures of the kinetic momenta remain unchanged during propagation.

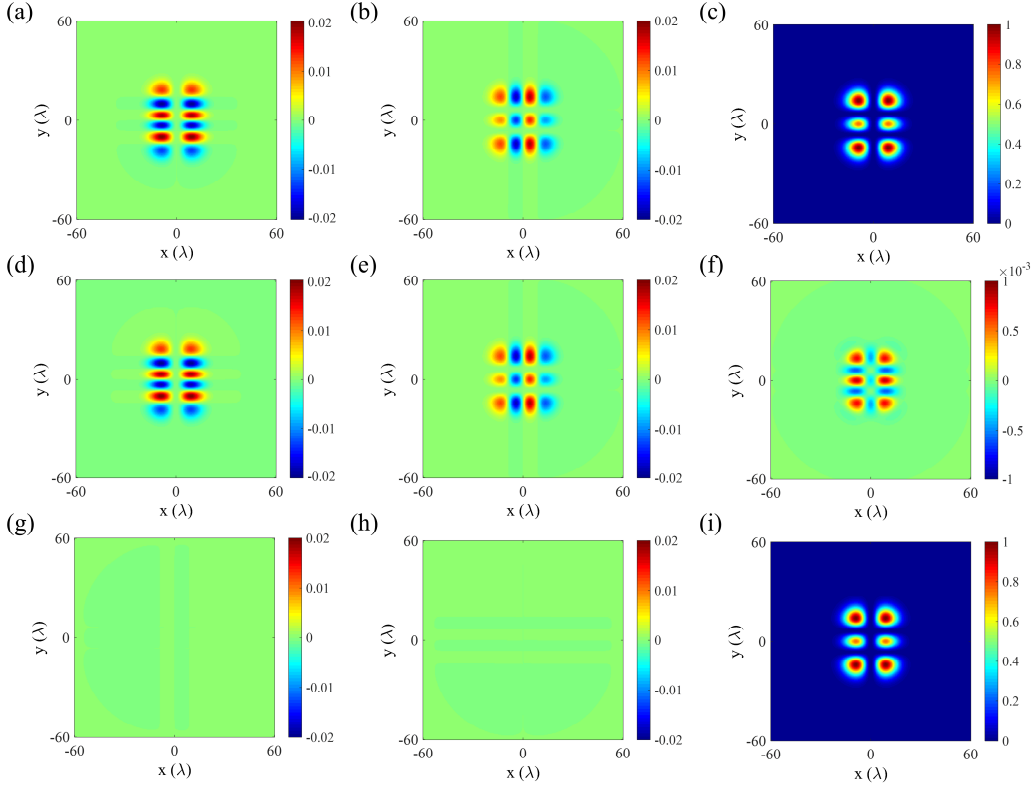

**Figure 2.** Spin properties of paraxial $HG_{12}$ beams in the focal plane. The total spin is three dimensional with: (a) *x*-component $S_x$, (b) *y*-component $S_y$, and (c), *z*-component $S_z$. The total spin decomposes into a T-spin that contributes to spin–momentum locking and a L-spin determined by the polarization ellipticities. The (d) *x*- and (e) *y*-component T-spins originate from the inhomogeneity of the kinetic momentum density and the (f) *z*-component T-spin relates to the Berry curvature. For the L-spin, the *x*- and *y*-components, (g) and (h) respectively, are considered as projections of the *z*-component (i) onto the horizontal axes. In the plane $z = 0.001\lambda$, both components are too small and can be ignored. The beam parameter settings are same as given in Figure 1.

To exhibit the spin property of the paraxial HG beam, the total SAM and the decomposed T-spin and L-spin given in Equation 1 were calculated,

$$\mathbf{S} = \frac{\varepsilon}{4\omega} \text{Im} \left\{ \begin{array}{l} \frac{1}{ik}\left[ +(\eta_x^*\eta_x + \eta_y^*\eta_y)\frac{\partial u_{HG}^* u_{HG}}{\partial y} - (\eta_x^*\eta_y - \eta_x\eta_y^*)\left( u_{HG}^* \frac{\partial u_{HG}}{\partial x} - u_{HG}\frac{\partial u_{HG}^*}{\partial x}\right) \right]\hat{\mathbf{x}} \\ \frac{1}{ik}\left[ -(\eta_x^*\eta_x + \eta_y^*\eta_y)\frac{\partial u_{HG}^* u_{HG}}{\partial x} - (\eta_x^*\eta_y - \eta_x\eta_y^*)\left( u_{HG}^* \frac{\partial u_{HG}}{\partial y} - u_{HG}\frac{\partial u_{HG}^*}{\partial y}\right) \right]\hat{\mathbf{y}} \\ 2\left[\eta_x^*\eta_y - \eta_y^*\eta_x\right] u_{HG}^* u_{HG} \hat{\mathbf{z}} \end{array} \right\}, \quad (6)$$

$$\mathbf{S}_T = \frac{1}{2k^2}\nabla\times\mathbf{p} = \frac{\varepsilon}{4\omega}\mathrm{Im}\left\{\begin{array}{c}\dfrac{1}{\mathrm{i}k}\left[+\left(\eta_x^*\eta_x+\eta_y^*\eta_y\right)\dfrac{\partial u_{\mathrm{HG}}^* u_{\mathrm{HG}}}{\partial y}\right]\hat{\mathbf{x}} \\ \dfrac{1}{\mathrm{i}k}\left[-\left(\eta_x\eta_x^*+\eta_y\eta_y^*\right)\dfrac{\partial u_{\mathrm{HG}}^* u_{\mathrm{HG}}}{\partial x}\right]\hat{\mathbf{y}} \\ \dfrac{1}{k^2}\left[\left(\eta_x^*\eta_x+\eta_y^*\eta_y\right)\left(\nabla u_{\mathrm{HG}}^*\times\nabla u_{\mathrm{HG}}\right)_z-\dfrac{1}{2}\left(\eta_x^*\eta_y-\eta_y^*\eta_x\right)\nabla_\perp^2\left(u_{\mathrm{HG}}^* u_{\mathrm{HG}}\right)\right]\hat{\mathbf{z}}\end{array}\right\}, \quad (7)$$

and

$$\mathbf{S}_L = \frac{\varepsilon}{4\omega}\mathrm{Im}\left\{\begin{array}{c}\dfrac{1}{\mathrm{i}k}\left[-\left(\eta_x^*\eta_y-\eta_x\eta_y^*\right)\left(u_{\mathrm{HG}}^*\dfrac{\partial u_{\mathrm{HG}}}{\partial x}-u_{\mathrm{HG}}\dfrac{\partial u_{\mathrm{HG}}^*}{\partial x}\right)\right]\hat{\mathbf{x}} \\ \dfrac{1}{\mathrm{i}k}\left[-\left(\eta_x^*\eta_y-\eta_x\eta_y^*\right)\left(u_{\mathrm{HG}}^*\dfrac{\partial u_{\mathrm{HG}}}{\partial y}-u_{\mathrm{HG}}\dfrac{\partial u_{\mathrm{HG}}^*}{\partial y}\right)\right]\hat{\mathbf{y}} \\ \left[\begin{array}{c}2\left(\eta_x^*\eta_y-\eta_y^*\eta_x\right)u_{\mathrm{HG}}^* u_{\mathrm{HG}} \\ -\dfrac{1}{k^2}\left(\eta_x^*\eta_x+\eta_y^*\eta_y\right)\left(\nabla u_{\mathrm{HG}}^*\times\nabla u_{\mathrm{HG}}\right)_z+\dfrac{1}{2k^2}\left(\eta_x^*\eta_y-\eta_y^*\eta_x\right)\nabla_\perp^2\left(u_{\mathrm{HG}}^* u_{\mathrm{HG}}\right)\end{array}\right]\hat{\mathbf{z}}\end{array}\right\}. \quad (8)$$

The total SAM given in Equation 6 are plotted in Figure 2a-c. The first terms of the horizontal SAMs are T-spins (Equation 7), which originate from the inhomogeneities of the EM fields and are proportional to the transverse gradients of the kinetic momenta in the paraxial approximation[17,41,42]; see Figure 2d-f. The other terms comprise the L-spin, which is determined by the polarization ellipticity of the EM field; see Figure 2g–i. The horizontal components of L-spin in Equation 8 can be understood further by considering the complex amplitude $u_{\mathrm{HG}}$ as an approximate superposition of plane waves $(\exp(\mathrm{i}(k_x x+k_y y)))$[125], and hence satisfy $\partial u_{\mathrm{HG}}/\partial x = \mathrm{i}k_x u_{\mathrm{HG}}$, $\partial u_{\mathrm{HG}}/\partial y = \mathrm{i}k_y u_{\mathrm{HG}}$ with $k_x/k$ and $k_y/k$ representing the horizontal directional vector components. Thus, these horizontal SAM components can be regarded as projections of the L-spin onto the horizontal axes when the EM wave is either converging or diverging. In this way, the horizontal SAM components are nearly zero at the focal plane; see Figure 2g–h.

Moreover, from the analysis above, the photons of paraxial HG modes possess spiral trajectories, leading to strong spin–orbit coupling, as evident in the $z$-component T-spin (Equation 9). Therein, the first term ($\mathrm{Im}\{\nabla u_{\mathrm{HG}}^*\times\nabla u_{\mathrm{HG}}\}\propto\nabla\times\mathbf{p}_o$) has a similar form to the Berry curvature of the optical potential[73,136], which is related to the evolution of the geometric phase in a paraxial optical system. The last term in Equation 7 originates from the vorticity of spin momentum $\mathbf{p}_s$, i.e., $\nabla\times\mathbf{p}_s = -\nabla^2\mathbf{S}/2 \approx -\hat{\mathbf{z}}\mathrm{Im}\{\eta_x^*\eta_y-\eta_y^*\eta_x\}\nabla_\perp^2(u_{\mathrm{HG}}^* u_{\mathrm{HG}})$ and $\mathbf{S}\approx\hat{\mathbf{z}}2\mathrm{Im}\{\eta_x^*\eta_y-\eta_y^*\eta_x\}u_{\mathrm{HG}}^* u_{\mathrm{HG}}$. Here, the transverse Laplace operator is $\nabla_\perp^2 = \partial^2/\partial x^2 + \partial^2/\partial y^2$. The physical meaning of this quantity can be further understood in a more general context. Employing the spin–orbit decomposition to calculate the vorticity of the kinetic momentum directly, one obtains[41]

$$\begin{aligned}\mathbf{S}_t &= \frac{1}{2k^2}\nabla\times\mathbf{p} = \frac{1}{2k^2}\nabla\times(\mathbf{p}_o+\mathbf{p}_s) = \frac{1}{2k^2}\left(-\frac{1}{\omega}\langle\nabla\psi|\times i|\nabla\psi\rangle-\frac{1}{2}\nabla^2\mathbf{S}\right) \\ &= \frac{\mathbf{S}}{2}-\frac{1}{8\omega^2}\mathrm{Re}\left\{\begin{array}{c}-(\nabla\otimes\mathbf{E}^*)\cdot\mathbf{H}-(\nabla\otimes\mathbf{E})^{\mathrm{T}}\cdot\mathbf{H}^* \\ +(\nabla\otimes\mathbf{H}^*)\cdot\mathbf{E}+(\nabla\otimes\mathbf{H})^{\mathrm{T}}\cdot\mathbf{E}^*\end{array}\right\}\end{aligned}, \quad (9)$$

where

$$\mathbf{r}_1 \otimes \mathbf{r}_2 = \begin{pmatrix} x_1 x_2 & y_1 x_2 & z_1 x_2 \\ x_1 y_2 & y_1 y_2 & z_1 y_2 \\ x_1 z_2 & y_1 z_2 & z_1 z_2 \end{pmatrix}. \qquad (10)$$

The T-spin given by the vorticity of kinetic momentum has a similar structure to the quantum 2-form[137] that generates the Berry phase associated with a circuit. Thus, the $z$-component of T-spin (Equation 7) can be regarded as the Berry curvature of the optical potential that originates from the spiral trajectories of photons. If polarization ellipticity is absent ($\eta_x = 0$ or $\eta_y = 0$ or the $\eta_x$ and $\eta_y$ are in-phase), the photons do not undergo spiral trajectories and thus the $z$-component T-spin is nearly zero, i.e., no Berry phase is generated in this instance. These Berry curvature-related terms are also included in the expression for L-spin (Equation 8) because they describe the evolution of the polarization ellipticities during propagation. Note that for paraxial optical systems the conditions imposed are $\partial^2/\partial x^2 \ll k\, \partial/\partial x \ll k^2$ and $\partial^2/\partial y^2 \ll k\, \partial/\partial y \ll k^2$ (as considered in Figure 2) and the integrals of the horizontal component of the L-spin and the Berry curvature-related term on the transverse plane are zero; one then concludes that the SAMs are conserved in the paraxial approximation, in agreement with references[138-141]. In summary, Equation 8, which represents the projection of the polarization ellipticity onto the 3D axes and includes the Berry curvature, is definitely the L-spin component.

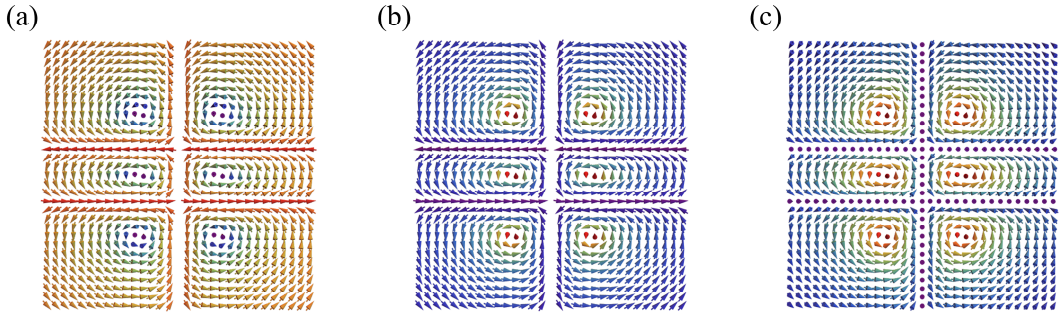

**Figure 3.** Momentum and spin textures of the paraxial HG$_{12}$ beam: (a) The vector diagram of kinetic momentum at the focal plane. The horizontal components contain multiple vortex structures, and thus the photons precess around the $z$-axis. (b) The normalized spin vector of the total spin. For each texture, the spin vector varies from the center "up" state to the boundary "horizontal" state, and thus the skyrmion number of each spin texture is $-1/2$ (half-skyrmion). (c) The normalized spin vector of the T-spin. The T-spin vectors vary from the center "up" state to the boundary "down" state and rotate in helical spirals, which is a manifestation of a Bloch-type skyrmion. The skyrmion number of each skyrmion-like spin texture is equal to $-1$. Skyrmion-like spin textures are formed in each transverse plane through spin–momentum locking of the T-spin and their topologies are maintained during propagation, free from the influence of the Gouy phase. Here, the topological number (skyrmion number) is calculated from $N = \iint_\sigma \mathbf{m} \cdot (\partial \mathbf{m}/\partial x \times \partial \mathbf{m}/\partial y)/4\pi$ [68], where $\mathbf{m} = \mathbf{S}/|\mathbf{S}|$ is the normalized spin vector.

In the confined optical field, a vortex structure of the kinetic momentum may result in a skyrmion-like spin texture[41,42,68]. Here, if there is a nonvanishing polarization ellipticity ($\text{Im}\{\eta_x^* \eta_y - \eta_y^* \eta_x\} \neq 0$), a similar vortex structure for the horizontal kinetic momentum is observed (Figure 3a). The vortex structure of the horizontal kinetic momentum causes the formation a skyrmion-like spin texture with T-spin through spin–momentum locking of the T-spin (Figure 3b, c). In contrast to the confined field case in which the spin vector of spin texture rotates in cycloidal spirals (Néel-type skyrmion), the spin texture of the paraxial HG beam possesses a Bloch-type configuration (Figure 3c), for which the spin vector rotates in helical spirals. Moreover, the skyrmion-like spin texture manifest spin–momentum locking and the kinetic momentum is maintained during propagation. Thus, the topology of the skyrmion-like spin texture is also unchanged during propagation and is unaffected by the evolution of the Gouy phase, in contrast to that of a paraxial skyrmionic beam[107]. Note that if there is a local disturbance (such as a nanoparticle) causing a kinetic momentum deflection, the topology of the skyrmion-like spin texture breaks down. The skyrmion number of each skyrmion-like spin texture is −1 when the polarization ellipticity $\text{Im}\{\eta_x^*\eta_y - \eta_y^*\eta_x\} > 0$.

**Spin-momentum property of Laguerre-Gaussian beams carrying phase singularities.**
Next, we consider a paraxial Laguerre-Gaussian (LG) beam[142] propagating along the $z$ axis in the cylindrical coordinates $(\rho, \varphi, z)$ with unit vector $(\hat{\boldsymbol{\rho}}, \hat{\boldsymbol{\varphi}}, \hat{\mathbf{z}})$. The electric and magnetic field components become

$$\mathbf{E}_{\text{LG}} = \left[\eta_x u_{\text{LG}}\hat{\mathbf{x}}, \eta_y u_{\text{LG}}\hat{\mathbf{y}}, +\frac{1}{ik}\left[\left(\eta_x\cos\varphi+\eta_y\sin\varphi\right)\frac{\partial u_{\text{LG}}}{\partial\rho}+\left(-\eta_x\sin\varphi+\eta_y\cos\varphi\right)\frac{i l u_{\text{LG}}}{\rho}\right]\hat{\mathbf{z}}\right]e^{-ikz}, \quad (11)$$

and

$$\mathbf{H}_{\text{LG}} = \frac{k}{\omega\mu}\left[\eta_y u_{\text{LG}}\hat{\mathbf{x}}, -\eta_x u_{\text{LG}}\hat{\mathbf{y}}, \frac{1}{ik}\left[\left(-\eta_x\sin\varphi+\eta_y\cos\varphi\right)\frac{\partial u_{\text{LG}}}{\partial\rho}-\left(\eta_x\cos\varphi+\eta_y\sin\varphi\right)\frac{i l u_{\text{LG}}}{\rho}\right]\hat{\mathbf{z}}\right]e^{-ikz}. \quad (12)$$

Here, $\eta_x$ and $\eta_y$ are arbitrary complex constants describing the relative strength and polarization ellipticity, respectively. We use these expressions to ensure the polarization state is homogeneous in the transverse plane. The electric field $\mathbf{E}_{\text{LG}}$ and magnetic field $\mathbf{H}_{\text{LG}}$ satisfy Gauss's law ($\nabla\cdot\mathbf{E}_{\text{LG}} = 0$ and $\nabla\cdot\mathbf{H}_{\text{LG}} = 0$) in the paraxial approximation. The complex amplitude $u_{\text{LG}}$ is given by

$$u_{\text{LG},pl} = \frac{w_0}{w(z)}\left[\frac{\sqrt{2}\rho}{w(z)}\right]^l L_p^l\left[\frac{2\rho^2}{w^2(z)}\right]\exp\left(-\frac{\rho^2}{w^2(z)}-i\frac{k\rho^2}{2R(z)}\right)\exp(il\varphi)\exp\left(-i(1+2p+|l|)\tan^{-1}\left(\frac{z}{z_R}\right)\right). \quad (13)$$

Here, $L_p^l(x)$ denotes the generalized Laguerre polynomial, $l$ the topological charge of vortex phase, and $\exp(-i(1+2p+|l|)\tan^{-1}(z/z_R))$ is the Gouy phase. From Equation 11 and 12, one obtains the kinetic momentum of the paraxial LG beam,

$$\mathbf{P} = \frac{1}{2}\frac{k}{\omega\mu}\text{Re}\begin{pmatrix}\frac{1}{ik}\left[+\left(\eta_\rho^*\eta_\rho+\eta_\varphi^*\eta_\varphi\right)u_{\text{LG}}^*\frac{\partial u_{\text{LG}}}{\partial\rho}+\eta_\rho^*\eta_\varphi\frac{1}{\rho}\frac{\partial u_{\text{LG}}^*u_{\text{LG}}}{\partial\varphi}\right]\hat{\boldsymbol{\rho}}\\ \frac{1}{ik}\left[+\left(\eta_\rho^*\eta_\rho+\eta_\varphi^*\eta_\varphi\right)u_{\text{LG}}^*\frac{1}{\rho}\frac{\partial u_{\text{LG}}}{\partial\varphi}-\eta_\rho^*\eta_\varphi\frac{\partial u_{\text{LG}}^*u_{\text{LG}}}{\partial\rho}\right]\hat{\boldsymbol{\varphi}}\\ \frac{1}{ik}\left[-ik\left(\eta_\rho^*\eta_\rho+\eta_\varphi^*\eta_\varphi\right)u_{\text{LG}}^*u_{\text{LG}}-0\right]\hat{\mathbf{z}}\end{pmatrix}, \quad (14)$$

which has a similar form to that of Equation 5. Here, we use substitutions $\eta_\rho = \eta_x\cos\varphi + \eta_y\sin\varphi$ and $\eta_\varphi = -\eta_x\sin\varphi + \eta_y\cos\varphi$. Thus, through energy conservation, the polarization ellipticities are $\eta_\rho^*\eta_\varphi - \eta_\varphi^*\eta_\rho = \eta_x^*\eta_y - \eta_y^*\eta_x$ and $\eta_\rho^*\eta_\rho + \eta_\varphi^*\eta_\varphi = \eta_x^*\eta_x + \eta_y^*\eta_y$. In this instance, $\eta_\rho$ and $\eta_\varphi$ are $\varphi$-dependent with $\partial\eta_\rho/\partial\varphi = \eta_\varphi$ and $\partial\eta_\varphi/\partial\varphi = -\eta_\rho$. The kinetic momentum of the paraxial LG$_{21}$ beam is found from Figure 4a, b. The first and second terms of Equation 14 are the canonical momentum and the Belinfante spin momentum, respectively, and $\partial u_{\text{LG}}^*u_{\text{LG}}/\partial\varphi = 0$. The paraxial LG beam carries phase singularities, which are associated with the orbital angular momenta (OAMs) of light[142]. In general, the photons of paraxial LG beam possess spiral trajectories, which derive from both the SAM and OAM of light.

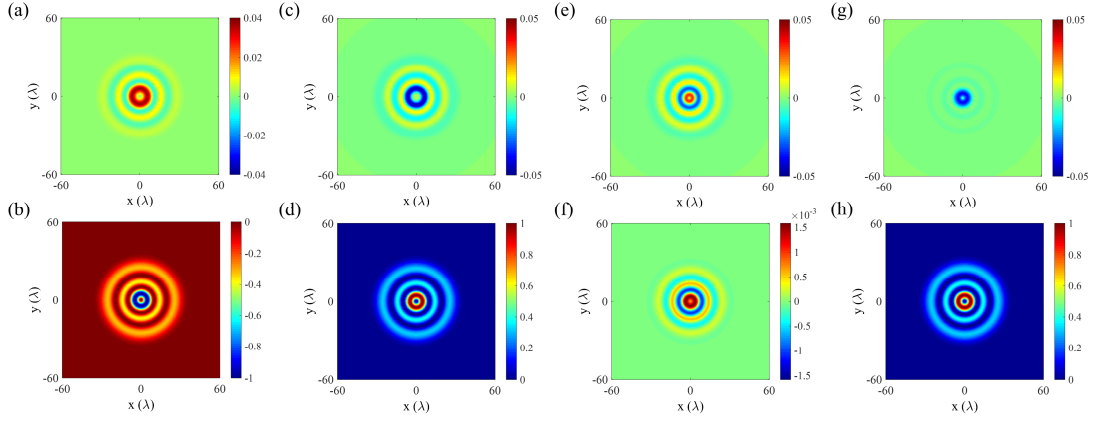

**Figure 4.** Spin–momentum properties of paraxial LG$_{21}$ beams in the focal plane $z = 0.001\lambda$: (a) azimuthal $p_\varphi$ and (b) z-component $p_z$ kinetic momenta result in spiral trajectories of the photons (the radial SAM vanishes in the focal plane); (c) azimuthal $S_\varphi$ and (d) z-component $S_z$ spin angular momenta decompose into T-spin and L-spin; (e) azimuthal T-spin stems from the radial gradient of z-component kinetic momentum density whereas (f) z-component T-spin is related to the Berry curvature; (g) azimuthal L-spins are considered projections of (h), the z-component L-spin onto the azimuthal direction. Here, $\eta_x = 1$, $\eta_y = 2i$, $w_0 = 8$ μm, $\lambda = 0.6328$ μm.

To exhibit the spin property of the paraxial LG beam, the total SAM and the decomposed T-spin and L-spin are expressed in the form

$$\mathbf{S} = \frac{\varepsilon}{4\omega} \text{Im} \begin{pmatrix} \frac{1}{ik}\left[0 - (\eta_\rho^*\eta_\varphi - \eta_\varphi^*\eta_\rho)\left(u_{LG}^* \frac{\partial u_{LG}}{\partial \rho} - u_{LG}\frac{\partial u_{LG}^*}{\partial \rho}\right)\right]\hat{\boldsymbol{\rho}} \\ \frac{1}{ik}\left[-(\eta_\rho^*\eta_\rho + \eta_\varphi^*\eta_\varphi)\frac{\partial u_{LG}^* u_{LG}}{\partial \rho} - (\eta_\rho^*\eta_\varphi - \eta_\varphi^*\eta_\rho)\left(u_{LG}^*\frac{1}{\rho}\frac{\partial u_{LG}}{\partial \varphi} - u_{LG}\frac{1}{\rho}\frac{\partial u_{LG}^*}{\partial \varphi}\right)\right]\hat{\boldsymbol{\varphi}} \\ 2[\eta_\rho^*\eta_\varphi - \eta_\varphi^*\eta_\rho]u_{LG}^* u_{LG}\hat{\mathbf{z}} \end{pmatrix}, \quad (15)$$

$$\mathbf{S}_T = \frac{1}{2k^2}\nabla \times \mathbf{p} = \frac{\varepsilon}{4\omega}\text{Im}\begin{pmatrix} 0\hat{\boldsymbol{\rho}} \\ \frac{1}{ik}\left[-(\eta_\rho^*\eta_\rho + \eta_\varphi^*\eta_\varphi)\frac{\partial u_{LG}^* u_{LG}}{\partial \rho}\right]\hat{\boldsymbol{\varphi}} \\ \frac{1}{k^2}\left[(\eta_\rho^*\eta_\rho + \eta_\varphi^*\eta_\varphi)(\nabla u_{LG}^* \times \nabla u_{LG})_z - \frac{1}{2}(\eta_\rho^*\eta_\varphi - \eta_\varphi^*\eta_\rho)\nabla_\perp^2(u_{LG}^* u_{LG})\right]\hat{\mathbf{z}} \end{pmatrix}, \quad (16)$$

and

$$\mathbf{S}_L = \frac{\varepsilon}{4\omega} \operatorname{Im} \left\{ \begin{array}{l} \dfrac{1}{ik}\left[-\left(\eta_\rho^* \eta_\varphi - \eta_\varphi^* \eta_\rho\right)\left(u_{LG}^* \dfrac{\partial u_{LG}}{\partial \rho} - u_{LG} \dfrac{\partial u_{LG}^*}{\partial \rho}\right)\right]\hat{\boldsymbol{\rho}} \\[6pt] \dfrac{1}{ik}\left[-\left(\eta_\rho^* \eta_\varphi - \eta_\varphi^* \eta_\rho\right)\left(u_{LG}^* \dfrac{1}{\rho}\dfrac{\partial u_{LG}}{\partial \varphi} - u_{LG} \dfrac{1}{\rho}\dfrac{\partial u_{LG}^*}{\partial \varphi}\right)\right]\hat{\boldsymbol{\varphi}} \\[6pt] \left[\begin{array}{l} +2\left(\eta_\rho^* \eta_\varphi - \eta_\varphi^* \eta_\rho\right)u_{LG}^* u_{LG} \\ -\dfrac{1}{k^2}\left[\left(\eta_\rho^* \eta_\rho + \eta_\varphi^* \eta_\varphi\right)\left(\nabla u_{LG}^* \times \nabla u_{LG}\right) + \dfrac{1}{2}\left(\eta_\rho^* \eta_\varphi - \eta_\varphi^* \eta_\rho\right)\nabla_\perp^2\left(u_{LG}^* u_{LG}\right)\right] \end{array}\right]\hat{\mathbf{z}} \end{array} \right\}. \quad (17)$$

The total spin components, Equation 15, are shown in Figure 4c, d. The first terms of the horizontal SAMs, which are proportional to the transverse gradients of the kinetic momenta, are T-spins, Equation 16, which are plotted in Figure 4e, f, whereas the other terms comprise the L-spin, Equation 17, which are plotted in Figure 4g, h. The horizontal L-spin components are regarded as projections of the total L-spin onto the horizontal axis as evident from Equation 8.

As mentioned above, for the paraxial LG modes, if the total angular momentum is nonzero, the photons exhibit spiral trajectories[142]. Interestingly, either the SAM or the OAM associated spiral trajectories generates a Berry phase. Thus, the z-component of the T-spin in Equation 16 and the last two terms in Equation 17 are related to the Berry curvature of the optical potential associated with the spiral trajectories of photons. If polarization ellipticity is absent ($\eta_x = 0$ or $\eta_y = 0$ or $\eta_x$ and $\eta_y$ are in-phase), the photons of the paraxial LG modes also possess spiral trajectories and a nonvanishing Berry curvature ($\operatorname{Im}\{\nabla u_{LG}^* \times \nabla u_{LG}\} \neq 0$) also exists. This is the primary distinction between the paraxial HG beam without optical singularities and the paraxial LG beam carrying phase singularities.

(a)          (b)          (c)

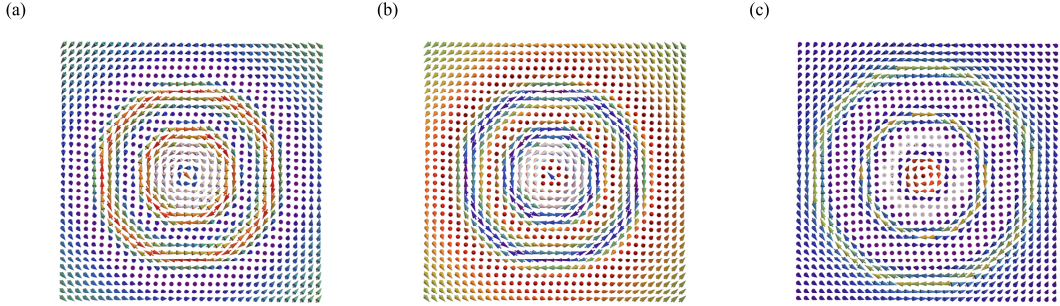

**Figure 5.** Momentum and spin vectors of the paraxial LG$_{21}$ beam: (a) The vector diagram of kinetic momentum at the focal plane. The horizontal components contain multiple vortex structures (along the radial direction), and thus the photons precess around the z-axis. (b) The normalized spin vector of the total spin. The spin vector varies from the center "up" state to boundary "horizontal" state in the radial direction, and thus the skyrmion number of each spin texture is −1/2 (half-skyrmion). (c) The normalized spin vector of the T-spin. For the center spin texture, the T-spin vectors vary from the center "up" state to boundary "down" state and rotate in helical spirals. The skyrmion number of this skyrmion-like spin texture is equal to −1. As the radius increases, the z-component T-spin is too small and the spin texture is unidentifiable.

In addition, for the paraxial system, the conditions $\partial^2/\partial\rho^2 \ll k\,\partial/\partial\rho \ll k^2$ and $1/k\rho \ll 1$ apply, the integrals of the horizontal component of L-spin and the Berry curvature-related term on the transverse plane are zero, and the SAMs are conserved in the paraxial approximation. Overall, Equation 17, which describes the 3D polarization ellipticity and Berry curvature, definitively defines the L-spin.

From Equation 14-17 and the above analysis, one finds that the introduction of phase singularities produces an additional contribution to spin–momentum locking, specifically, an OAM-associated Berry curvature in the T-spin. This OAM-associated Berry curvature was also reported in references[126,127]. In addition, the kinetic momentum of paraxial LG mode contains the vortex structure (Figure 5a), and hence the skyrmionlike spin texture constructed by T-spin appears at the center (Figure 5b, c).

**Spin-momentum property of Bessel-Gaussian beams carrying polarization singularities.**
Finally, we consider a paraxial Bessel-Gaussian (BG) mode[143] carrying polarization singularities propagating along the $z$ axis in cylindrical coordinates ($\rho$, $\varphi$, $z$) with unit vectors ($\hat{\rho}$, $\hat{\varphi}$, $\hat{z}$). The electric and magnetic field components become

$$\mathbf{E}_{BG} = \left[\cos\Phi u_{BG}\hat{\rho}, \sin\Phi u_{BG}\hat{\varphi}, \frac{1}{ik}\left[\frac{1}{\rho}\frac{\partial\rho u_{BG}}{\partial\rho} + \frac{(n-1)u_{BG}}{\rho}\right]\cos\Phi\hat{z}\right]^T e^{-ikz}, \quad (18)$$

and

$$\mathbf{H}_{BG} = \frac{k}{\omega\mu}\left[\sin\Phi u_{BG}\hat{\rho}, -\cos\Phi u_{BG}\hat{\varphi}, \frac{1}{ik}\left[\frac{1}{\rho}\frac{\partial\rho u_{BG}}{\partial\rho} + \frac{(n-1)u_{BG}}{\rho}\right]\sin\Phi\hat{z}\right]e^{-ikz}. \quad (19)$$

Here, we use the substitution $\Phi = [(n-1)\varphi + \varphi_0]$ with $\varphi_0$ the initial angle of the polarization state. When $n = 0$, the electric/magnetic field distributions prescribe a paraxial Gaussian beam carrying a homogeneous polarization. For $n \neq 0$, the polarization state is inhomogeneous in the transverse propagating plane and depends on the azimuthal coordinate $\varphi$. There is a polarization singularity at the center of the BG beam. The electric field $\mathbf{E}_{BG}$ and magnetic field $\mathbf{H}_{BG}$ also satisfy Gauss' law ($\nabla\cdot\mathbf{E}_{BG} = 0$ and $\nabla\cdot\mathbf{H}_{BG} = 0$) in the paraxial approximation. The complex amplitude $u_{BG}$ is given by[143]

$$u_{BG} = \frac{1}{1+iz/z_R}\exp\left(-\frac{\rho^2/w_0^2}{1+iz/z_R}\right)\exp\left[-\frac{i\beta^2 z/(2k)}{1+iz/z_R}\right]J_n\left(\frac{\beta\rho}{1+iz/z_R}\right), \quad (20)$$

where $\beta$ denotes a constant that determines the beam profile ($\beta$ is positive for this study), and $J_n$ the $n$-order Bessel function of the first kind. From the expressions for the electric and magnetic fields of these BG modes, one obtains a kinetic momentum of

$$\mathbf{p} = \frac{1}{2}\frac{k}{\omega\mu}\text{Re}\left[\frac{1}{ik}\left[+u_{BG}^*\frac{\partial u_{BG}}{\partial\rho}\right]\hat{\rho}, 0\hat{\varphi}, \frac{1}{ik}\left[-iku_{BG}^*u_{BG}\right]\hat{z}\right]. \quad (21)$$

In the case, there is only canonical momentum, and Belinfante spin momentum is zero approximately (Figure 6a, e). Moreover, the photons do not possess spiral trajectories ($p_\varphi = 0$) and no Berry phase is generated. The total SAM and its T-spin and L-spin components are

$$\mathbf{S} = \frac{\varepsilon}{4\omega}\text{Im}\left[0\hat{\rho}, \frac{1}{ik}\left[-\frac{\partial u_{BG}^*u_{BG}}{\partial\rho} - 2n\frac{u_{BG}^*u_{BG}}{\rho}\right]\hat{\varphi}, 0\hat{z}\right]^T, \quad (22)$$

$$\mathbf{S}_T = \frac{1}{2k^2}\nabla\times\mathbf{p} = \frac{\varepsilon}{4\omega}\text{Im}\left[0\hat{\rho}, \frac{1}{ik}\left[-\frac{\partial u_{BG}^*u_{BG}}{\partial\rho}\right]\hat{\varphi}, 0\hat{z}\right]^T, \quad (23)$$

and

$$\mathbf{S}_L = \frac{\varepsilon}{4\omega}\text{Im}\left[0\hat{\rho}, \frac{1}{ik}\left[-2\frac{nu_{BG}^*u_{BG}}{\rho}\right]\hat{\varphi}, 0\hat{z}\right]^T. \quad (24)$$

In Equation 22, one finds that the total SAM (Figure 6b, f) contains two terms: the azimuthal SAMs, which are proportional to the transverse gradients of kinetic momenta, are T-spins (Figure 6c, g) as given in Equation 23, whereas the other terms are determined by the polarization topological charge. When the polarization topological charge $n = 0$, the polarization state is homogeneous; the result is consistent with that of Equation 8, i.e., no L-spin exists. However, if $n \neq 0$, the directional vector of this SAM component is determined by the inherent property of the mode, namely, the polarization topological charge (Figure 6d, h). Thus, this SAM component in Equation 24 can be regarded as an L-spin, although it is localized in the transverse plane of the kinetic momentum. This extraordinary L-spin, which does not generate spin–momentum locking but is a $\mathbb{Z}_4$ topological invariant, is interesting and was discovered quite recently[44,57]. Obviously, no skyrmion-like spin textures arise because the $z$-component SAM vanishes. In summary, the introduction of polarization singularities in the paraxial beam creates an extraordinary L-spin but no skyrmion-like spin textures.

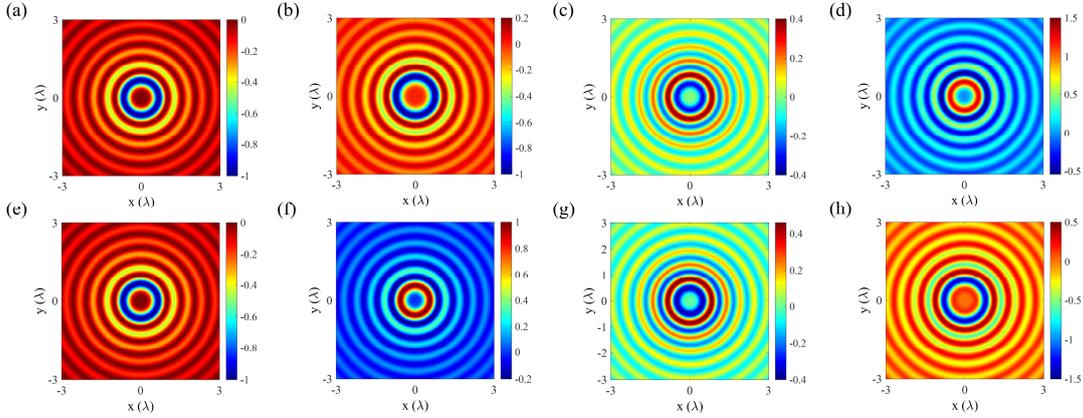

**Figure 6.** Spin-momentum properties of paraxial BG beams in the focal plane $z = 0$: (a) $z$-component kinetic momentum $p_z$, (b) azimuthal optical spin $S_\varphi$, (c) optical T-spin, (d) optical L-spin for the +3-order paraxial BG beams, (e) $z$-component kinetic momentum $p_z$, (f) azimuthal optical spin $S_\varphi$, (g) optical T-spin, and (h) optical L-spin for the −3-order paraxial BG beam. The other components are zero. From (a) and (e), the kinetic momenta are the same, and thus the T-spin are also the same; see (c) and (g). However, the L-spin is inverted when the polarization topological charge switches between +3 and −3. This spin, which is independent of the vorticity of kinetic momentum but depends on the polarization singularity, is the L-spin (and is a $\mathbb{Z}_4$ topological invariant). Here, $w_0 = 8$ μm, $\lambda = 0.6328$ μm.

**III Conclusion**

To summarize, we presented a unified methodology to describe the spin–momentum properties in the paraxial optical systems. The theory uncovers the underlying physical difference between T-spin and L-spin. Moreover, the decomposition of T-spin is consistent with the Helmholtz decomposition theory[144] and can be generalized to other classical wave fields[44,145]. We investigated the influence of optical singularities on the spin–momentum properties of paraxial optical beams. For the HG beam without optical singularities, skyrmion-like spin textures may arise from the T-spins when the beam elliptically polarized. Then, for a LG beam with a phase singularity, the OAM property of the phase singularity results in a Berry curvature term in the T-spin, and thus a skyrmion-like spin texture arises from the T-spins when the total angular momentum of the beam is nonzero. Moreover, for the BG mode with a polarization singularity, no skyrmion-like spin texture is formed. Nevertheless, we discovered a SAM component, the direction of which is not determined by the kinetic momentum but rather the polarization topological charge of vector vortex beam. This SAM component is a $\mathbb{Z}_4$ topological invariant and should be considered as an extraordinary L-spin, although its vector is perpendicular to the canonical momentum/wavevector. The findings present a field theory to describe the spin–momentum properties of paraxial systems having potential in formalizing a spin-based theory to understand wave–matter interactions in classical wave fields and motivating the exploration of novel applications in chiral manipulation and interdisciplinary research[146–151].